\documentclass[prd,preprint]{revtex4}

\usepackage{stmaryrd}
\usepackage{mathrsfs}
\usepackage{amsmath}
\usepackage{latexsym}
\usepackage{amsfonts}
\usepackage{graphicx}
\usepackage{subfigure}
\begin{document}
\title{A Multiple Step-like Spectrum of Primordial Perturbation}

\author{Jie Liu}
\email{liujie10b@mails.gucas.ac.cn}
\author{Yun-Song Piao}
\email{yspiao@gucas.ac.cn}

\affiliation{ College of Physical Sciences, Graduate University of
Chinese Academy of Sciences, Beijing 100049, China}

\begin{abstract}

We show that if the inflaton effective potential has multiple
discontinuous points in its first derivative, the spectrum of
primordial perturbation will be multiple step-like. We give a
general analysis by applying a simple model. In principle, as long
as the height of step is low enough, the result of spectrum will
be consistent with observations.

\end{abstract}
\maketitle

\section{Introduction}
Inflation\cite{Guth},\cite{acc2},\cite{acc3} is a stage of an
exponential expansion, which brings the current observable universe.
During the period of inflation, the quantum fluctuations inside the
horizon would be able to extend out the horizon and become the
primordial perturbations responsible for the structure formation of
observable universe, see Ref.\cite{acc5}. The primary advantages of
inflation and the accumulating observational data make the inflation
scenario become a leading candidate of the primordial universe.

The primordial perturbations obtained during inflation is
approximately scale invariant and Gaussian, all of which have been
confirmed by experiments to some extent. In some inflationary
models, inflaton potentials has a step, and the steplike change in
the potential will result in an universal oscillation in the
spectrum of primordial perturbations
\cite{Adams},\cite{Leach},\cite{acc9},\cite{acc10},\cite{Hamann},\cite{Chen07},\cite{Chen08},\cite{acc12},\cite{Bean08},\cite{acc13},\cite{acc18},\cite{Liu11},\cite{Frederico},\cite{David1},\cite{David2}.
In other inflationary models, the oscillation of primordial spectrum
can be also generated, e.g. inflaton potential with a small
oscillation \cite{Wang02},\cite{Liddle08},\cite{Fl10},\cite{KO11},
the field decay during inflation \cite{BBD}, or the particles
production during inflation \cite{Sasaki08},\cite{L09},\cite{B09},
the bounce before inflation \cite{acc14},\cite{Cai},\cite{M08}, the
heterotic M-theory \cite{Ashoorioon:2006wc}. Interestingly, a burst
of oscillations in the primordial spectrum seems to provide a better
fit to the CMB angular power spectrum\cite{acc6,acc7}. However,
given the improvements in data quality anticipated with coming
observations, it is significant to ask whether arbitrary features in
the primordial power spectrum can be acquired during inflation.
Therefore, it is worth to search other possibilities of the features
of the primordial power spectrum.

As has been pointed in \cite{acc16},\cite{Starobinsky97},\cite{HS},
if the inflationary potential has a discontinuity in its first
derivative at certain point, there will a step-like behavior in its
power spectrum. As long as the height of this step is low enough,
the inflationary power spectrum obtained can be still agreement with
the observations. Recently, the inflationary models based on the
string landscape have been intensively studied. It might be possible
that there are multiple discontinuous points in the first derivative
of inflationary effective potential. In this paper, we analytically
and numerically show that this might imply a multiple step-like
spectrum of primordial perturbation. We will give a general analysis
by applying a simple model, and therefore the conclusion is somewhat
universal.

The plan of this work is as follows. The power spectrum of
perturbations is calculated in section II. The numerical results
of this power spectrum are showed in section III. In section IV,
we discuss the applications of results.

\section{The calculations of power spectrum}

We begin with the inflaton effective potential with a discontinuity
in its first derivative at some points. Hereafter, we call such
points as the discontinuous points for convenience. In the paper, we
only discuss the simplest case of the effective potential, i.e. the
linear potential except the discontinuous point. We define
$V,_{\varphi}=d_{j}$ for $\varphi_{j}< \varphi <\varphi_{j-1}$,
where $d_{j}$ is the slope of the potential in the corresponding
region and is constant, and $d_{j}>0$, $d_{j}\neq d_{j-1}$. In each
region, $V,_{\varphi\varphi}=0$, thus the slow roll conditions can
be satisfied for small $d_{j}$. However, at the discontinuous points
($\varphi_j$),
\begin{equation}
V,_{\varphi\varphi} =  \sum_{j\,=1}^n {\cal
D}_{j-1,j}\,\delta(\varphi-\varphi_{j}),
 \label{1}
\end{equation}
where ${\cal D}_{j-1,j}=d_{j-1}-d_j$, and $n$ is the number of the
discontinuous points. The sum of delta functions is rewritten as
\begin{equation} V,_{\varphi\varphi} \simeq {3H \over
a}\sum_{j\,=1}^n {{\cal D}_{j-1,j}\over d_j}\,\delta(\eta-\eta_{j}),
 \label{2}\end{equation}
where $3H\dot{\varphi}\simeq -V,_{\varphi}$ and Eq.(\ref{1}) are
applied. Here, the delta functions are obviously induced by the
discontinuity of first derivative of the effective potential, which
will inevitably appear in the perturbation equation via
$V,_{\varphi\varphi}$. We will check the effect of such
discontinuous points on the primordial perturbation in the
following.

The equation of motion for the scalar perturbation $\delta\varphi$
during inflation in $k$ space is given by, after defining
$v=a\delta\varphi$, and $x=k\eta$,
\begin{equation}
\frac{d^2v}{dx^2}+[1-\frac{2}{x^{2}}+\sum_{j\,=1}^n\frac{\sigma_{j}}{x}\delta(x-x_{j})]v=0,
\label{3}
\end{equation}
where $\sigma_{j}={ 3{\cal D}_{j-1,j} \over d_{j}}\ll 1$ is required
by the observations, and ${2}/{x^{2}}$ is given by
$a^{\prime\prime}/a\simeq 2/\eta^2$.
 In principle, although the
field, due to some mechanisms, suddenly experiences a change of its
slope, the change should be not large. $\sigma_{j}$ is critical for
the shape of the power spectrum, which will be found.

The solutions of Eq.(\ref{3}) are,
\begin{equation}
 v_{j} =
a_{j}(i+\frac{1}{x})e^{-ix}+b_{j}(-i+\frac{1}{x})e^{ix},
\label{vi}
\end{equation} where $a_j$ and $b_j$ are dependent of
$k$. In general, initially the perturbation mode is deep inside
the horizon, which implies $a_{0}\sim {1\over \sqrt{2k}},b_{0}=0$,
i.e.
\begin{equation}
v_{0} = a_{0}(i+\frac{1}{x})e^{-ix}.
\end{equation}
By applying the junction condition
\begin{equation}
v_{j-1}(x_{j})=v_{j}(x_{j}),~[\frac{dv}{dx}]_{x_{j}}=-\,\frac{v(x_{j})}{x_{j}}\sigma_{j},
\end{equation}
we have the Bogoliubov coefficients
\begin{eqnarray}
a_{j} =[1+\frac{\sigma_{j}}{2ix_{j}}(1+\frac{1}{x^{2}_{j}})]a_{j-1}+\frac{\sigma_{j}}{2ix_{j}}(\frac{1}{x_{j}}-i)^2e^{2ix_{j}}b_{j-1},\nonumber\\
b_{j}
=-\frac{\sigma_{j}}{2ix_{j}}(i+\frac{1}{x_{j}})^2e^{-2ix_{j}}a_{j-1}+[1-\frac{\sigma_{j}}{2ix_{j}}(1+\frac{1}{x^{2}_{j}})]b_{j-1},
\label{7}
\end{eqnarray}
where $x_{j}={k}{\eta_{j}}$. This is a set of recursive equations,
by which $a_j$ and $b_j$ can be related to $a_0$ and $b_0$. We are
interested in the perturbations on large scale, i.e. evaluated at
$x=k\eta\rightarrow 0$. Thus from Eq.(\ref{vi}), we have
\begin{eqnarray}
v  = a_{n}(i+\frac{1}{x})e^{-ix}+b_{n}(-i+\frac{1}{x})e^{ix} \simeq
\frac{a_{n}+b_{n}}{x}\,\,\, for\,\,\, x\rightarrow 0.
 \label{8}
\end{eqnarray}
 We get
\begin{eqnarray}
|v|^{2}=\frac{1}{|x|^{2}}[|M_{n}|^{2}(|a_{n-1}|^{2}+|b_{n-1}|^{2})+2Re(M_{n}^{2}a_{n-1}b_{n-1}^{*})],
 \label{9}
\end{eqnarray}
\begin{eqnarray}
M_{n}=1+\frac{\sigma_{n}}{x_{n}}(\frac{sinx_{n}}{x_{n}}-cosx_{n})(\frac{1}{x_{n}}+i)e^{-ix_{n}}.
 \label{10}
\end{eqnarray}
Therefore the power spectrum of $\delta\varphi$ is
\begin{equation}
\mathcal {P}_{\delta\varphi}(k)=\frac{k^3}{2\pi^2}|{v\over a}|^{2}
 \label{11},
 \end{equation}
 and the spectral index is
 \begin{equation}
n_{\delta\varphi}-1=\frac{d\ln\mathcal
{P}_{\delta\varphi}(k)}{d\ln k}
 \label{12}.
  \end{equation}
where $v_k$ is given by Eq.(\ref{9}), which is a set of recursive
equations and is difficultly to be solved for the case $n\gg 1$.
Thus we will give the numerical results of spectrum in following
section.

\section{ The Numerical Results}

When the effective potential of field only includes a discontinuous
point, for Eqs.(\ref{7}), we read
\begin{eqnarray}
a_{1}& = &a_{0}[1+\frac{\sigma_{1}}{2ix_{1}}(1+\frac{1}{x^{2}_{1}})],\nonumber\\
b_{1}& =
&-a_{0}\frac{\sigma_{1}}{2ix_{1}}(i+\frac{1}{x_{1}})^2e^{-2ix_{1}}.
 \label{13}
\end{eqnarray}
Thus with Eq.(\ref{8}), we have
\begin{equation}
v\approx\frac{a_{0}}{x}[1+\frac{\sigma_{1}}{x_{1}}(\frac{sinx_{1}}{x_{1}}-cosx_{1})(\frac{1}{x_{1}}+i)e^{-ix_{1}}].
 \label{14}
\end{equation}
The result is that obtained in Ref.\cite{acc6}. The power spectrum
generally has a jump, which is leaded by the discontinuous point,
see also \cite{Gong05}.

In this paper, we will see that for an effective potential with
multiple discontinuous points, the power spectrum will has a
multiple step-like behavior.

When the potential have two discontinuous points during the field
rolling down, with Eq.(\ref{7}) and Eq.(\ref{13}), the Bogoliubov
coefficients are
\begin{eqnarray}
a_{2}&=&[1+\frac{\sigma_{2}}{2ix_{2}}(1+\frac{1}{x^{2}_{2}})]a_{1}+\frac{\sigma_{2}}{2ix_{2}}(-i+\frac{1}{x_{2}})^{2}e^{2ix_{2}}b_{1}~~~~~~~~~\nonumber\\
&=&
a_{0}[1+\frac{\sigma_{1}}{2ix_{1}}(1+\frac{1}{x^{2}_{1}})][1+\frac{\sigma_{2}}{2ix_{2}}(1+\frac{1}{x^{2}_{2}})]-\frac{a_{0}\sigma_{1}\sigma_{2}}{(2i)^2x_{1}x_{2}}(i+\frac{1}{x_{1}})^{2}e^{2ix}(-i+\frac{1}{x_{2}})^{2}e^{2i(x_{2}-x_{1})},\nonumber\\
b_{2}&=&-\frac{\sigma_{2}}{2ix_{2}}(i+\frac{1}{x_{2}})^2e^{-2ix_{2}}a_{1}+[1-\frac{\sigma_{2}}{2ix_{2}}(1+\frac{1}{x^{2}_{2}})]b_{1}~~~~~~~\nonumber\\
&=&-\frac{a_{0}\sigma_{2}}{2ix_{2}}(i+\frac{1}{x_{2}})^2[1+\frac{\sigma_{1}}{2ix_{1}}(1+\frac{1}{x^{2}_{1}})]e^{-2ix_{2}}-\frac{a_{0}\sigma_{1}}{2ix_{1}}(i+\frac{1}{x_{1}})^{2}[1-\frac{\sigma_{2}}{2ix_{2}}(1+\frac{1}{x^{2}_{2}})]e^{-2ix_{1}}.
 \label{15}
\end{eqnarray}
So
\begin{eqnarray}
v\approx\frac{a_{0}}{x}\{{1+\frac{\sigma_{1}}{x_{1}}(i+\frac{1}{x_{1}})(\frac{sinx_{1}}{x_{1}}-cosx_{1})e^{-ix_{1}}}+
\frac{\sigma_{2}}{x_{2}}(i+\frac{1}{x_{2}})(\frac{sinx_{2}}{x_{2}}-cosx_{2})e^{-ix_{2}}~~~~~~~~~~~~\nonumber\\
+\frac{\sigma_{1}\sigma_{2}}{x_{1}x_{2}}(i+\frac{1}{x_{1}})e^{-ix_{1}}(\frac{sinx_{2}}{x_{2}}-cosx_{2})[(1+\frac{1}{x_{1}x_{2}})cos(x_{1}-x_{2})+(\frac{1}{x_{2}}-\frac{1}{x_{2}})(sin(x_{1}-x_{2})]\}.~~~~
\end{eqnarray}

\begin{figure}
\subfigure[]{ \label{step20:subfig:step22:subfig:step21}
\begin{minipage}[b]{0.3\textwidth}
\centering
\includegraphics[width=5.1cm, height=3cm]{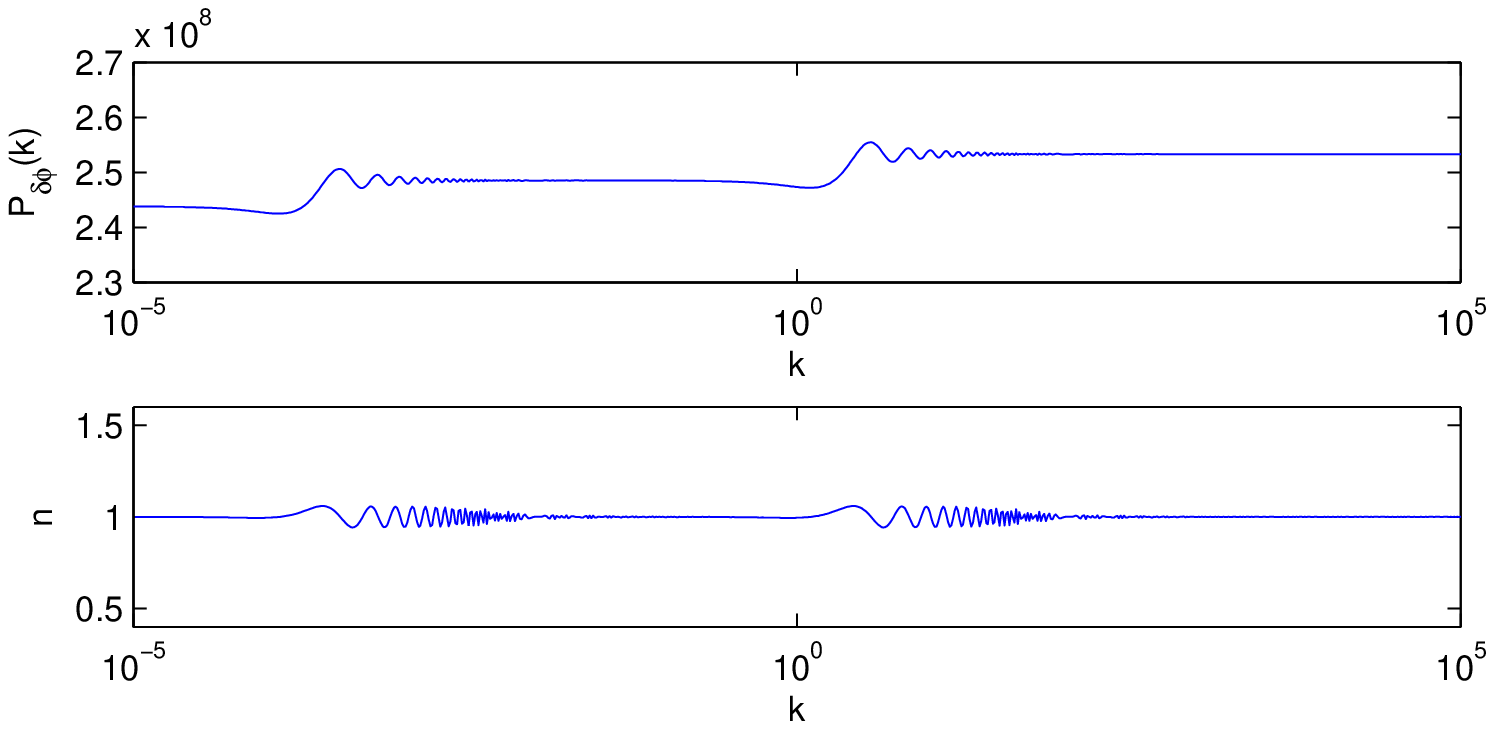}
\end{minipage}}%
\subfigure[]{ \label{step20:subfig:step22:subfig:step21}
\begin{minipage}[b]{0.3\textwidth}
\centering
\includegraphics[width=5.0cm, height=3cm]{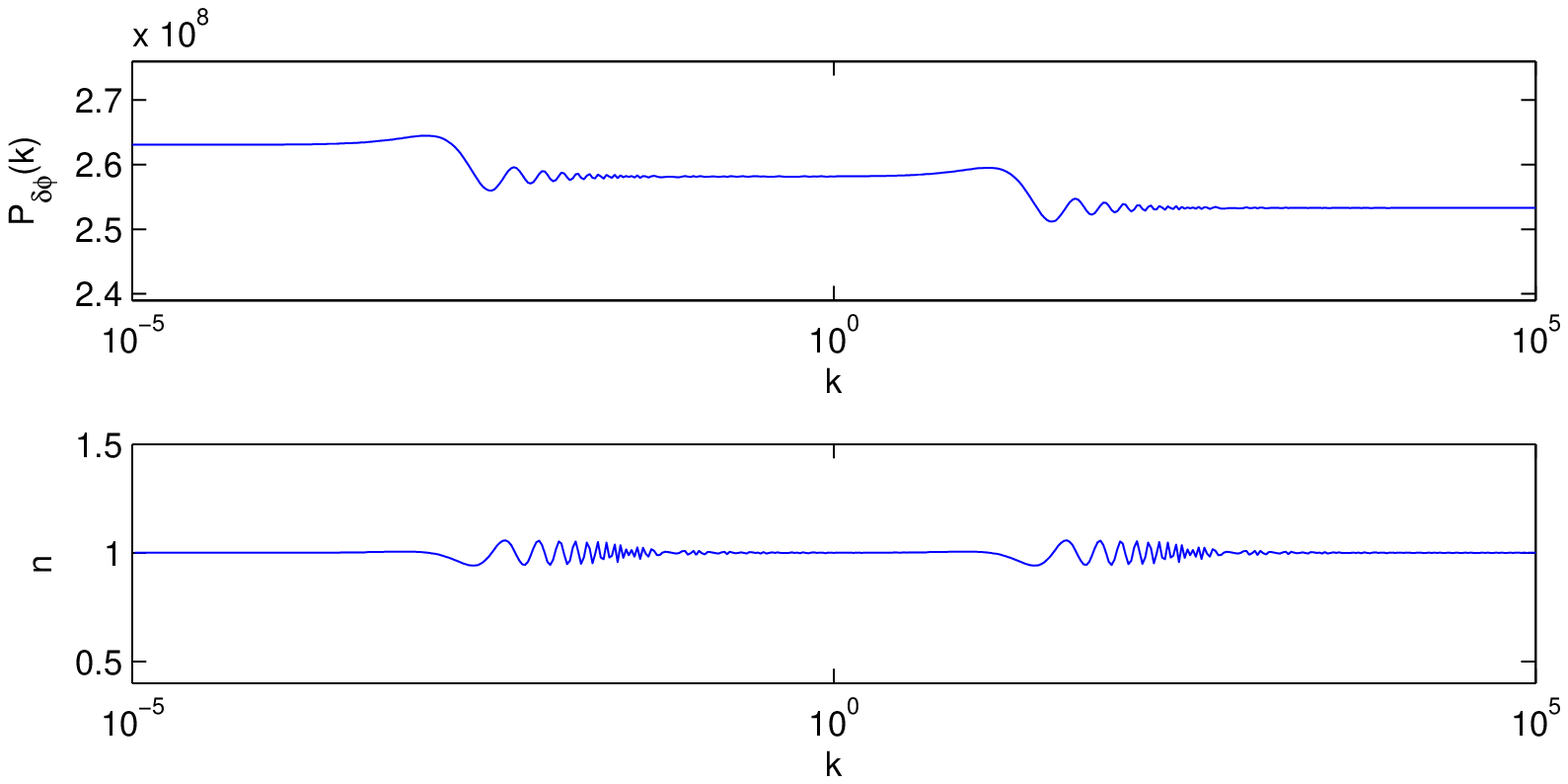}
\end{minipage}}
\subfigure[]{ \label{step20:subfig:step22:subfig:step21}
\begin{minipage}[b]{0.3\textwidth}
\centering
\includegraphics[width=5.0cm, height=3cm]{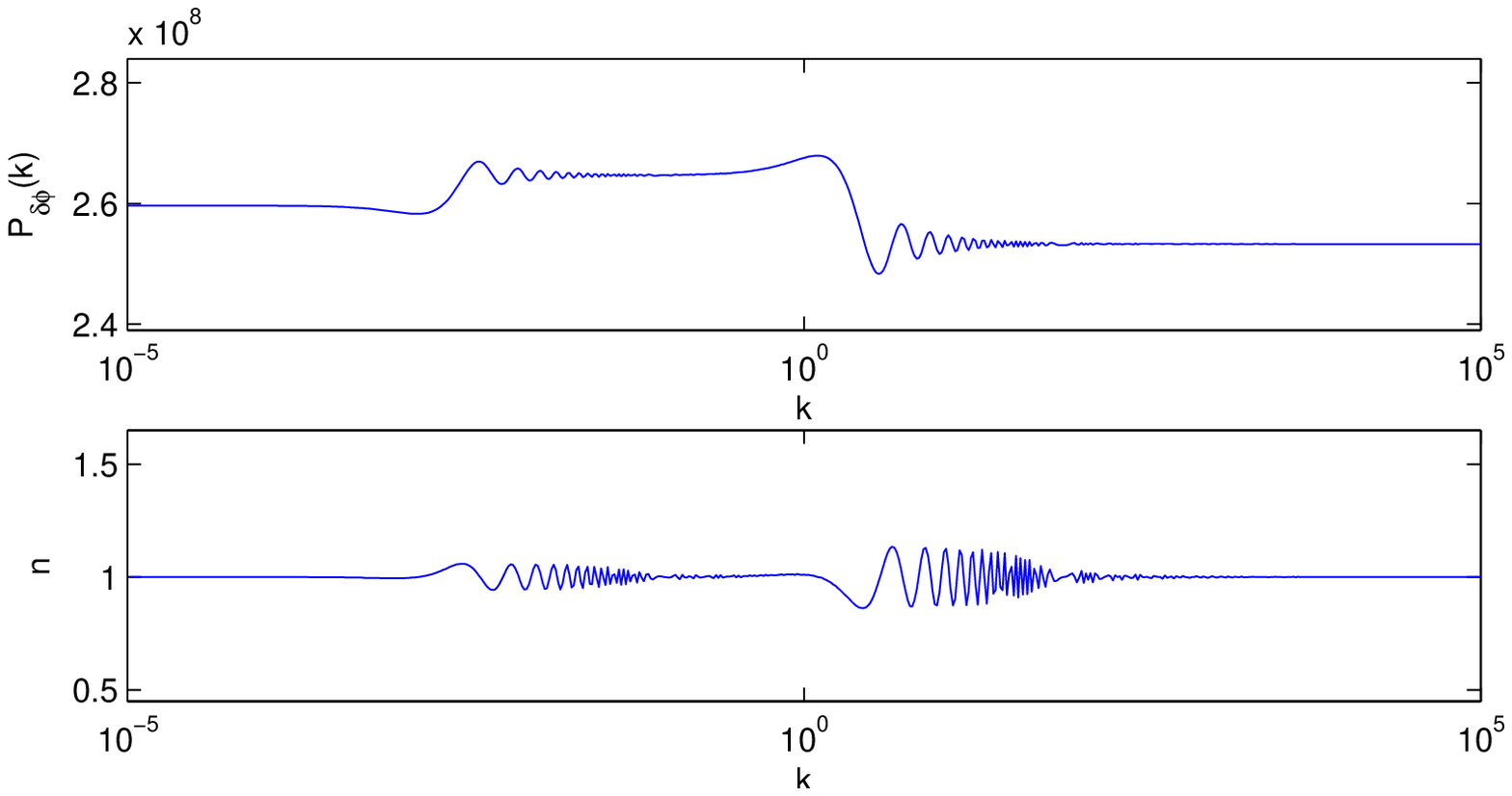}
\end{minipage}}
 \caption{${\cal P}_{\delta\varphi}~ and~ n_{\delta\varphi}$ with two steps:
(a) $\sigma_{1}=-1/35,~\sigma_{2}=-1/35$; (b)
$\sigma_{1}=1/35,~\sigma_{2}=1/35$; (c)
$\sigma_{1}=-1/35,~\sigma_{2}=1/15$. Where the unit of $k$ is
$Mpc^{-1}$. }
 \label{fig1}
\end{figure}

We plot numerically ${\cal P}_{\delta\varphi}$ and
$n_{\delta\varphi}$ in Fig.\ref{fig1}. Fig.\ref{fig1} shows ${\cal
P}_{\delta\varphi}$ has two steps and there is generally
oscillations around each step. However, the oscillations only occur
in a narrow region and the amplitude of oscillation rapidly decay.
For $n_{\delta\varphi}$, there is also vibration in the
corresponding narrow region and the amplitude is small and rapidly
decay. In the Fig.\ref{fig1}(a) and (b), the height of both steps
are same because of $|\sigma_{1}|=|\sigma_{2}|$, and the power
spectrum is `up' when $\sigma < 0$ but `down' when $\sigma>0$. In
the Fig.\ref{fig1}(c), the power spectrum is firstly `up' and then
`down' because of $\sigma_{1}<0, ~\sigma_{2}>0$ and the level of
downward jump is larger than upward because of
$|\sigma_{1}|<|\sigma_{2}|$, so the power spectrum of the large
scale to little scale perturbations `jump' down because of
$\sigma_{1}+\sigma_{2}>0$.

Further, although the spectrum experiences many time jumps and the
value of $\sigma$ is random, the values of the power spectrum on
smallest and largest $k$ may be same when \begin{equation}
\sum_{j=1}^{n}\sigma_{j}=0.\label{17}\end{equation} This can be
observed as follows.

\begin{figure}
\subfigure[]{ \label{step80:subfig:step81}
\begin{minipage}[b]{0.5\textwidth}
\centering
\includegraphics[width=7.5cm, height=3cm]{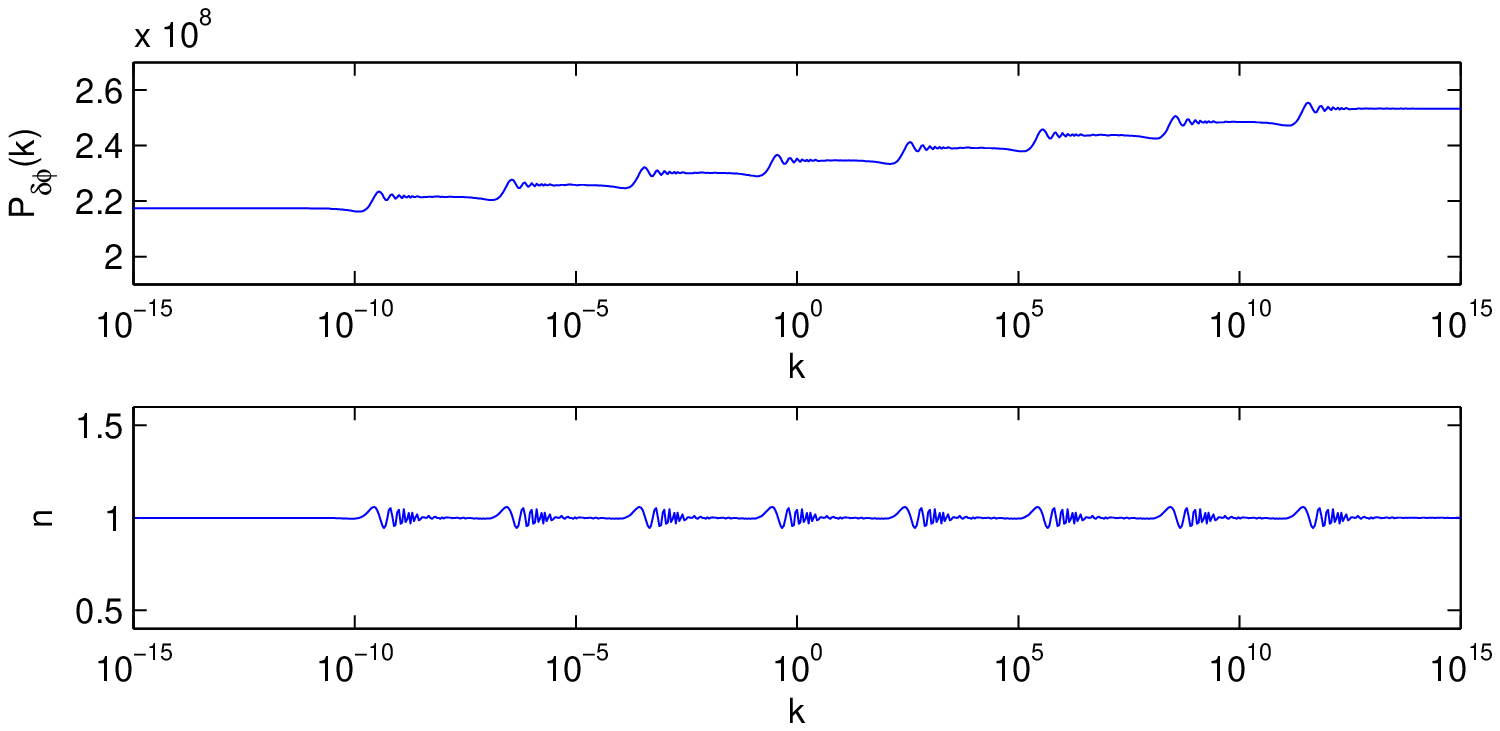}
\end{minipage}}%
\subfigure[]{ \label{step80:subfig:step81}
\begin{minipage}[b]{0.5\textwidth}
\centering
\includegraphics[width=7.5cm, height=3cm]{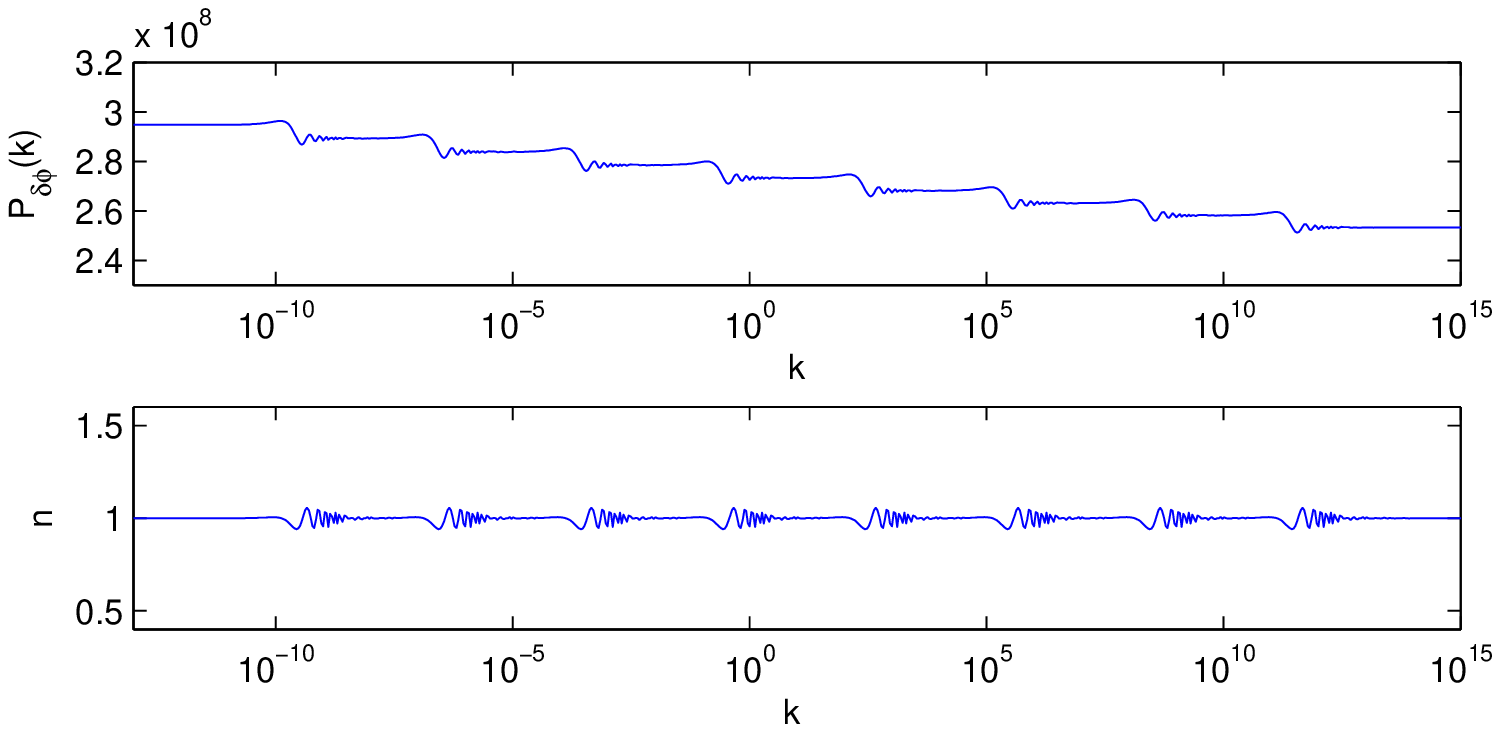}
\end{minipage}}
 \caption{${\cal P}_{\delta\varphi}~ and ~n_{\delta\varphi}$ with eight steps:
(a) $\sigma_{j}=-1/35$;~ (b) $\sigma_{j}=1/35, ~j=1,2,...,8.$ Where
the unit of $k$ is $Mpc^{-1}$.
 }
 \label{fig2}
\end{figure}

\begin{figure}
\subfigure[]{ \label{step82:subfig:step83}
\begin{minipage}[b]{0.5\textwidth}
\centering
\includegraphics[width=7.5cm, height=3cm]{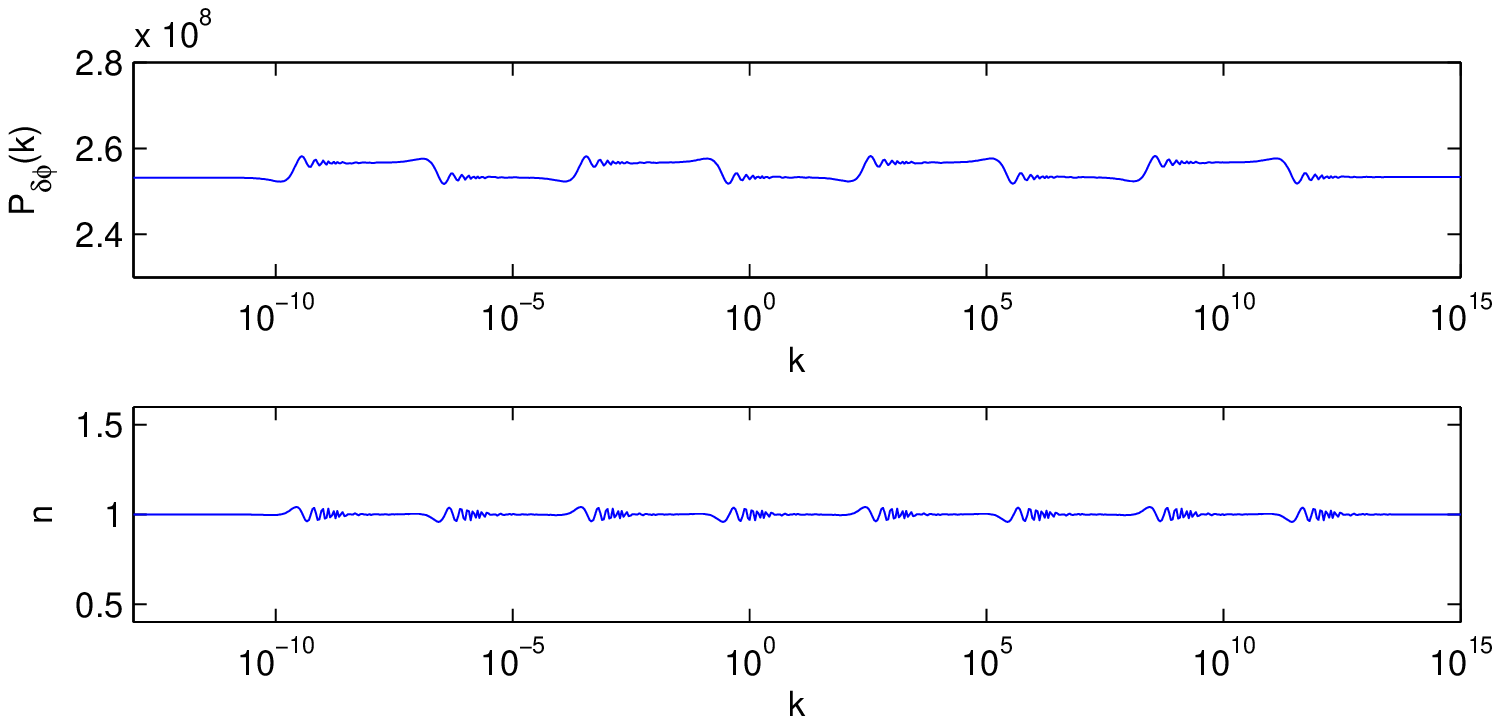}
\end{minipage}}%
\subfigure[]{ \label{step82:subfig:step83}
\begin{minipage}[b]{0.5\textwidth}
\centering
\includegraphics[width=7.5cm, height=3cm]{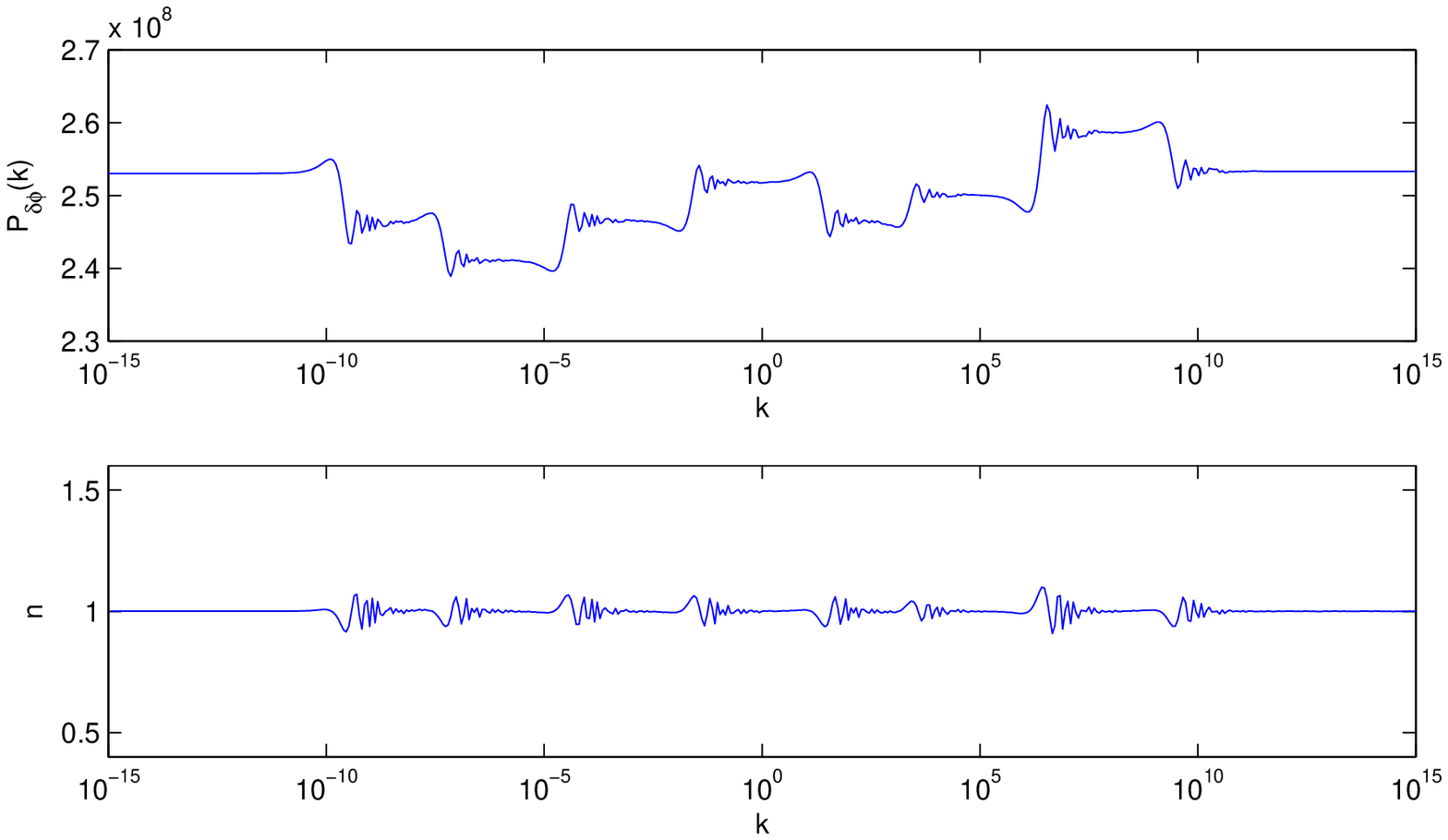}
\end{minipage}}
 \caption{${\cal P}_{\delta\varphi}~ and~ n_{\delta\varphi}$ with eight steps:
 (a) $\sigma_{j}=(-)^{j}1/35,~j=1,2,...,8$; (b) $\sigma_{1}=1/24, \sigma_{2}=1/32,
\sigma_{3}=-1/30,\sigma_{4}=-1/32,
\sigma_{5}=1/32,\sigma_{6}=-1/48,\sigma_{7}=-1/20, \sigma_{8}=1/32$.
Where the unit of $k$ is $Mpc^{-1}$. }
 \label{fig3}
\end{figure}

Though the simplest case in Fig.\ref{fig1} has grasped the points of
multiple step-like behavior, it is still interesting to check the
detailed case for the effective potential with $n>2$ discontinuous
points. We solve Eq.(\ref{11}) numerically for the equation
including eight delta functions, and the results are plotted in
Figs.\ref{fig2} and \ref{fig3}. In general, the number of the
`steps' equals the number of the discontinuous points in the
effective potential, and the shape of the spectrum depends on the
parameter $\sigma_{i}$. In the Fig.\ref{fig2}(a), the power spectrum
jumps `up' eight times with the same value of $\sigma_{i}<0$, and is
alike with an upstairs. According to the value of
$\sum_{j=1}^{8}\sigma_{j}$, the power spectrum  show a large `jump'
from the largest scale to smallest scale perturbations, while it is
inverse in the Fig.\ref{fig2}(b). The Fig.\ref{fig3}(a) shows a
`Great Wall' and  Fig.\ref{fig3}(b) shows a random wall, and the
values of both the power spectrum on smallest and largest $k$ are
same with Eq.(\ref{17}).

\section{Discussion}

We have showed that if the inflaton effective potential has multiple
discontinuous points in the first derivative, the spectrum of
primordial perturbation will be multiple step-like. We have given a
general analysis by applying a simple model. Actually, For a
nonlinear effective potential with  multiple discontinuous points in
the first derivative, the conclusion is the same as what we discuss
in the paper, only if in the region $\varphi_{j}< \varphi
<\varphi_{j-1}$, the potential can generate the flat power spectrum.
In principle, as long as the height of step is low enough, the
result of spectrum will be consistent with observations. However, it
can be noticed that one or several significant steps might bring
unexpected feature in CMB, which is interesting to further study.

The oscillation in primordial spectrum has appeared in many
inflationary models,but a multiple step-like spectrum shows a new
feature. Though there is generally oscillations around each step,
the oscillations only occur in a narrow region and the amplitude of
oscillation rapidly decay; the multiple step exert little influence
on the whole spectrum except several unusual points, especially in
Fig.\ref{fig3}. Therefore the result given here might have
interesting applications in future.

Theoretically, the smooth inflation potential is perfect, but it is
probable that, in reality, due to some reasons or mechanisms, the
potential is rude or discontinuous in its first derivative, and the
smooth only is local. On the landscape, a potential with valleys,
hills, and steep as well as some shallow regions is more generic.
The effective potential may be multiple discontinuous points in the
first derivative, so the spectrum of primordial perturbation will be
similar to the discussion in the paper.

In Ref.\cite{acc18} for meandering inflation, it is argued that
during inflation, inflaton would meander in a complicated
multi-dimensional potential, thus during different phases the slope
of potential is different, and its change is random. In this model,
the effective potential is similar to that used here, thus it might
be possible that one could find multiple step-like feature in its
power spectrum.

In Ref.\cite{acc19}, the effect of the decay of fields during
inflation, which leads to a staggered inflation \cite{BBD}, on the
primordial spectrum is computed. Whenever a field decays, its
associated potential energy is transferred into radiation, causing
a jump in the equation of state parameter. Thus there are the
discrete steps in the power spectrum. The jump in the equation of
state parameter in certain sense might be equivalent to the change
of the slope of effective potential, thus in this sense the result
is slightly similar to that here.

Recent CMB observations seems to favor the primordial power
spectrum with features. We will back to a detailed compare of a
multiple step-like power spectrum with CMB observation, which will
certainly impose the constrains on the number of steps and its
height.

\textbf{Acknowledgments} This work is supported in part by NSFC
under Grant No:10775180, 11075205, and 11005165, in part by the
Scientific Research Fund of GUCAS(NO:055101BM03), in part by
National Basic Research Program of China, No:2010CB832804.

\end{document}